\documentstyle[12pt]{article}

\makeatletter
\ifcase\@ptsize
  \font\tenmsy=msbm10
  \font\sevenmsy=msbm7
  \font\fivemsy=msbm5
\or
  \font\tenmsy=msbm10 scaled \magstephalf
  \font\sevenmsy=msbm8
  \font\fivemsy=msbm6
\or
  \font\tenmsy=msbm10 scaled \magstep1
  \font\sevenmsy=msbm8
  \font\fivemsy=msbm6
\fi
\newfam\msyfam
\textfont\msyfam=\tenmsy  
\scriptfont\msyfam=\sevenmsy
\scriptscriptfont\msyfam=\fivemsy
\def\Bbb{\ifmmode\let\next\Bbb@\else
\def\next{\errmessage{Use \string\Bbb\space only in math mode}}\fi\next}
\def\Bbb@#1{{\Bbb@@{#1}}}
\def\Bbb@@#1{\fam\msyfam#1}
\newfam\euffam
\font\sixeuf=eufm6
\font\eighteuf=eufm8
\font\twelveeuf=eufm10 scaled\magstep1
\textfont\euffam=\twelveeuf
\scriptfont\euffam=\eighteuf
\scriptscriptfont\euffam=\sixeuf
\def\euf{\fam\euffam\twelveeuf}
\makeatother

\newcommand{\BZ}{{\Bbb{Z}}}
\newcommand{\Bid}{1\!{\rm l}}

\newcommand{\myboldmath}{\boldmath}
\def\pn{\par\noindent}
\def\pano{\par\noindent}

\def\w{{\cal W}}
\def\del{\partial}

\def\eps{\varepsilon}
\def\lam{\lambda}
\def\be{\begin{equation}}
\def\ee{\end{equation}}
\def\ba{\begin{array}}
\def\ea{\end{array}}
\def\bea{\begin{eqnarray}}
\def\eea{\end{eqnarray}}
\def\bean{\begin{eqnarray*}}
\def\eean{\end{eqnarray*}}
\def\bl{\begin{list}{}{}}

\def\ds{\displaystyle}

\newcommand{\reseteqn}{\setcounter{equation}{0}}
\newcommand{\mysection}{\reseteqn\section}
\renewcommand{\thefootnote}{\fnsymbol{footnote}}

\if@twoside
\else
\fi
\begin{document}
  \pagestyle{empty}
  \begin{raggedleft}
CSIC-IMAFF-47-1995\\
hep-th/9605152\\
October 1995\\
  \end{raggedleft}
  $\phantom{x}$ 
  {\LARGE\bf
  \begin{center}
Fusion \& Tensoring\\ of\\
Conformal Field Theory\\ and\\
Composite Fermion Picture\\ of\\ 
Fractional Quantum Hall Effect  
  \end{center}
  }\par
  \vfill
  \begin{center}
$\phantom{X}$\\
{\Large Michael A.I.~Flohr\footnote[1]{address after December $1^{{\rm st}}$,
1995: School of Natural Sciences,
Institute for Advanced Study, Olden Lane, Princeton, NJ 08540, USA, 
email: {\tt flohr@sns.ias.edu}}}\\
{\em Instituto de Matem{\'a}ticas y F{\'\i}sica Fundamental, C.S.I.C.,\\
Serrano 123, E-28006 Madrid, Spain\\
email: {\tt iffflohr@roca.csic.es}}
  \end{center}\par
  \vfill
  \begin{abstract}
  \noindent 
  We propose a new way for describing the transition between two quantum
  Hall effect states with different filling factors within the framework
  of rational conformal field theory. Using a particular class of non-unitary
  theories, we explicitly recover Jain's picture of attaching flux quanta
  by the fusion rules of primary fields. Filling higher Landau levels of
  composite fermions can be described by taking tensor products of conformal
  theories. The usual projection to the lowest Landau level corresponds then
  to a simple coset of these tensor products with several U(1)-theories
  divided out. These two operations -- the fusion map and the tensor map --
  can explain the Jain series and all other observed fractions as 
  exceptional cases. Within our scheme of transitions we naturally find a 
  field with the experimentally observed universal critical exponent $7/3$. 
  \end{abstract}
  \vfill
  \newpage
%
%
  \setcounter{page}{1}
  \pagestyle{plain}
  \renewcommand{\thefootnote}{\roman{footnote}}
  \mysection{Introduction}
  \pn 
It is well known that the wave functions of quantum Hall effect (QHE) 
states can be recovered from correlation functions of chiral conformal
field theories (CFTs) \cite{CMM91,Fub91,MR91}. We first review this
principle with the slight generalization of an arbitrary background charge.
Let $\phi(z)$ be a chiral scalar free field. Then $j=\del\phi$ obeys an
{\euf u}(1)-Kac-Moody algebra
  \be
    [j_m,j_n] = n\delta_{m+n,0}\,,
  \ee
which is known to describe the chiral edge waves, i.e.\ the energy gapless
excitations of the QHE states. Introduce a Virasoro field by 
the Sugawara construction
  \be 
    L(z)\ =\ :jj:(z) + \sqrt{2}\alpha_0\del_z j(z)\,,
  \ee
where $:\ldots:$ means normal ordering. The Virasoro algebra has then
central charge $c = 1 - 24\alpha_0^2$ with $2\alpha_0$ the background charge.
The local primary fields are constructed from vertex operators $\psi_{\alpha}$
of conformal weight $h(\alpha) = \alpha^2 - 2\alpha\alpha_0$, explicitly
  \be\label{eq:Vop}
    \psi_{\alpha}=\exp\left(-\sum_{n>0}\sqrt{2}\alpha j_n\frac{z^n}{n}\right)
                  \exp\left(-\sum_{n<0}\sqrt{2}\alpha j_n\frac{z^n}{n}\right)
                  c(\alpha)z^{-\sqrt{2}\alpha\alpha_0}\,,
  \ee
where $c(\alpha)$ commutes with all $j_n, n\neq 0$ and maps highest-weight
states into highest-weight states. 
The vertex operators are well defined if the charges 
$\alpha$ are restricted to the set $\alpha_{r,s} =\frac{1}{2}(1-r)\alpha_-
+\frac{1}{2}(1-s)\alpha_+$ with $r,s\in\BZ$ and $\alpha_{\pm} =
\alpha_0\pm\sqrt{\alpha_0^2+1}$. A more careful study shows that under 
particular circumstances CFTs can be consistent and well defined local 
primary fields can exist even with certain rational values of $r,s$.
This precisely happens \cite{Flo93} if $\frac{1}{4}\alpha_0^2\in\BZ_+$.
In the following we
denote the local primary fields and their chiral parts as
$\Phi_{(n,n'|\bar n,\bar n')}(z,\bar z) \equiv
\Phi_{h_{n,n'},\bar h_{\bar n,\bar n'}}(z,\bar z) =
\Phi_{n,n'}(z)\otimes\Phi_{\bar n,\bar n'}(\bar z)$, 
where the indices may be rational.
  \par
Consider a generic correlation function of chiral vertex operators 
(\ref{eq:Vop}) on the plane. One has the well known result
  \begin{equation}\label{eq:vev}
    \langle\Omega_{2\alpha_0}^*,\psi_{\alpha_1}(z_1)\ldots\psi_{\alpha_N}(z_N)
    \Omega_0\rangle = \prod_{i<j}(z_i-z_j)^{2\alpha_i\alpha_j}\,,
  \end{equation}
if $|z_1|>\ldots>|z_N|$ and $\sum_{i=1}^N\alpha_{i} = 0$, where
$\Omega_{\alpha}$ denotes the ground state to the superselection sector of 
charge $\alpha$. 
To reproduce the non-holomorphic parts of e.g.\ the Laughlin wave
functions \cite{Lau83}
  \begin{equation}\label{eq:laughlin}
    \Psi_{\nu=\frac{1}{2p+1}} = 
    \prod_{i<j}(z_i-z_j)^{2p+1}\exp\left(-\frac{1}{2}\sum_i|z_i|^2\right)\,,
  \end{equation}
we insert a term $\exp(-i\alpha\int d^2z'\bar{\rho}\phi(z'))$ 
into the correlator (\ref{eq:vev}), where $\phi$ 
is again the free field and $\bar{\rho}$ is an averaged density
$(\pi\alpha^2)^{-1}$. If one integrates this term over a disk of area
$2\pi\alpha^2N$, then the real part correctly yields the desired
exponential term for $N$ electrons, while the imaginary part contributes
a singular phase. The latter can be eliminated by an also singular 
gauge transformation corresponding to the uniform external magnetic field
\cite{MR91}. In the following we will often neglect the exponential term
and absorb the external magnetic field in $\Omega_{2\alpha_0}^*(N)$, 
since the integral also modifies the background charge.
  \par
Unfortunately we must bother now the reader by sketching briefly a very
particular set of {\em non-unitary\/} rational conformal field theories 
(RCFTs) -- the ones
with $\frac{1}{4}\alpha_0^2\in\BZ_+$ -- since they are at the very heart
of the paper.
  \par
Thus, let $\eps\in\{0,1\}$ and let $\min\lam$ denote the smallest 
representative of $\lam\in\BZ/m\BZ$. Then for every 
$k$ there exist two RCFTs, one with
  \be\ba{rcll}
    c & = & 1 - 24k\,, & k\in\BZ_+/2 \\
    {\ds h_{\frac{\lam}{2k+2\eps},(-)^{\eps}\eps\frac{\lam}{2k+2\eps}}} &
        = & {\ds \left[\left(\frac{\min\lam}{2k+2\eps}\right)^2 - 1\right]k
            + \eps\left(\frac{\min\lam}{2k+2\eps}\right)^2}\,, &
                      \lam\in\BZ/(k+\eps)\BZ \\
    h_{1,1} & = & 0
  \ea\ee
which has the extended chiral symmetry algebra $\w(2,3k)$, and its
$\BZ_2$ orbifold
  \be\ba{rcll}
    c & = & 1 - 24k\,, & k\in\BZ_+/4 \\
    {\ds h_{\frac{\lam}{4k+4\eps},(-)^{\eps}\frac{\lam}{4k+4\eps}}} &
        = & {\ds \left[\left(\frac{\min\lam}{4k+4\eps}\right)^2 - 1\right]k
            + \eps\left(\frac{\min\lam}{4k+4\eps}\right)^2}\,, &
                      \lam\in\BZ/(4k+4\eps)\BZ \\
    h_{1,1} & = & 0 \\
    h_{2,2} & = & 3k
  \ea\ee
with chiral symmetry algebra $\w(2,8k)$. Here $h_{r,s}$ denotes the
Virasoro highest weight analogous to the Virasoro highest weights 
of degenerate models to generic central charge $c = 1 - 24\alpha_0^2$
given by $h_{r,s} = \frac{1}{4}\left((r\alpha_-+s\alpha_+)^2
- (\alpha_-+\alpha_+)^2\right)$. Note that in contrast to the generic 
degenerate
Virasoro model $s = \pm r$ and $r$ is not restricted to integers only.
For further details see \cite{BFKNRV91,Flo93}. All these RCFTs have
effective central charge $c_{{\em eff}} = c - 24h_{{\em min}} = 1$.
  \par
The finitely many highest weight representations are highest weight
representations with respect to the extended symmetry algebra. The
characters which are infinite sums of Virasoro characters can be
expressed in terms of Jacobi-Riemann $\Theta$-functions divided by the
usual Dedekind $\eta$-function. For example the vacuum character of the
$\w(2,3k)$ theories is given by
\bea
  \chi^{}_{{\em vac}}(\tau) & = & {\ds\sum_{n\in\BZ_+}\chi_{h_{n,n}}^{\em
                               Vir}(\tau)\ =\ 
                               q^{\frac{1-c}{24}}\sum_{n\in\BZ_+}
                               \frac{q^{h_{n,n}}-q^{h_{n,-n}}}{\eta(\tau)}} \\
                         & = & {\ds \frac{1}{2\eta(\tau)}\left(
                               \Theta_{0,k}(\tau) - \Theta_{0,k+1}(\tau)
                               \right)}\,,
\eea
where $\eta(\tau) = q^{1/24}\prod_{n=1}^{\infty}(1-q^n)$,
$\Theta_{\lam,k}(\tau) = \sum_{n\in\BZ}q^{(2kn+\lam)^2/4k}$, and
$q = e^{2\pi i\tau}$. The other characters can be obtained by the modular
transformation $S:\tau\mapsto-\frac{1}{\tau}$. Details including
$S$ and $T$ matrix and fusion rules can be found in \cite{Flo93}.
  \par
It follows that these RCFTs consist of two sectors, each a Gaussian theory,
which are twisted by the vacuum representation and the $h_{{\em min}}$
representation, common to both sectors. The both Gaussian sectors
have different boundary conditions, i.e.\ different compactification
radii $2R^2 = p/q$ and $2R'^2 = p'/q'$. We have the following condition
in order to get consistent RCFTs: $p'q'-pq = \Delta$ for
$k\in\BZ_++\frac{1}{\Delta}$, $\Delta\in\{1,2,4\}$.
Note the similarity to the unimodular equation defining ${\rm SL}(2,\BZ)$.
Therefore, we suggestively
denote these RCFTs with ${\euf C}\left[{p'\ p\atop q\ q'}\right]$.
Note that the twist of two Gaussian sectors has the property that fusion
of two fields in one sector yields fields in the other sector, i.e.\ that
boundary conditions and thus statistics phases are changed.
%
%
  \par
  \mysection{Laughlin states {\myboldmath $\nu=1/(2p+1)$}}
  \pn 
Let us first concentrate on QHE states with filling factor $\nu=1/(2p+1)$,
i.e.\ the Laughlin states.
We now make the following specific choice of the fermionic non-unitary 
RCFTs introduced above with background charges $\alpha_0 = \sqrt{(2p+1)/2}$, 
i.e.\ $c = 1 - 24\frac{2p+1}{2}$ and $\Delta=2$.
We work with the full rational conformal field theory (RCFT),
but with $\BZ_2$ frustrated boundary conditions in one direction, and 
introduce the chirality constraint by hand. The frustrated partition function
  \bea
    Z_{1,0} &=&
    \chi_0\chi_{h_{{\em min}}}^* + \chi_{h_{{\em min}}}\chi_0^*\nonumber\\  
    &+&
    \sum_{\lambda\in\BZ_{2p+1}-\{0\}}{\textstyle\frac{1}{2}}
      (\chi_{\lambda,{\em even}}\chi_{\lambda,{\em odd}}^* + c.c.) + 
    \sum_{\lambda\in\BZ_{2p+3}-\{0\}}{\textstyle\frac{1}{2}}
      (\chi_{\lambda,{\em even}}\chi_{\lambda,{\em odd}}^* + c.c.)
  \eea      
is modular invariant under the subgroup $\Gamma(2)\subset\Gamma=
{\rm PSL}(2,\BZ)$ generated by $T^2:\tau\mapsto\tau+2$ and $ST^2S:\tau\mapsto
\frac{\tau}{\tau+2}$. The characters have been split into even and odd
contributions modulo the fermion number $(-)^F$.
In particular, the groundstate of this frustrated
RCFT is twofold degenerated (one for each direction of the external magnetic
field) and is created by the local primary fields $\Phi_{(0,0|1,1)} = 
\psi_{\alpha_0}(z)\otimes\Bid$ and $\Phi_{(1,1|0,0)}$ respective, thus is
{\em chiral}. The conformal weight is $(h,\bar h) = (h_{{\em min}},0) = 
(-\frac{2p+1}{2},0)$ or vice versa. The $N$-point correlator evaluates to
  \begin{equation}\label{eq:myvev}
    \langle\Omega_{\sqrt{2(2p+1)}}^*(N),\prod_{i=1}^N\Phi_{(0,0|1,1)}(z_i,
    \bar z_i)\Omega_0\rangle = 
    \prod_{i<j}(z_i-z_j)^{2p+1}\exp\left(-\frac{1}{2}\sum_i|z_i|^2\right)\,,
  \end{equation}
hence nicely reproduces the Laughlin wave functions (\ref{eq:laughlin}).
It is remarkable that the Laughlin QHE state now appears as {\em groundstate}
of a certain RCFT. But this is not the only remarkable fact. As we will show
now, attaching of flux quanta has a beautiful realization within these
RCFTs given by fusion of primary fields.  
  \par
First we identify the vertex operator which describes
a single magnetic flux quantum by its anyonic statistics as quasi-particle
excitation. While the composite fermions are described by 
$\Phi_{(0,0|1,1)}(z,\bar z)$, the flux quantum is realized by the field
$\Phi_{(\frac{2p}{2p+1},\frac{2p}{2p+1}|\frac{2p}{2p+1},\frac{2p}{2p+1})}
(w,\bar w)$, whose conformal dimension is $(h,\bar h)=(-\frac{4p+1}{4p+2}, 
-\frac{4p+1}{4p+2})$. Inserting $M$ such flux quanta into the correlator
(\ref{eq:myvev}), we obtain
  \begin{eqnarray}\label{eq:anyon}\lefteqn{%
    \langle\Omega_{\sqrt{2(2p+1)}}^*(N)
    \Omega_{\sqrt{2/(2p+1)}}^*(M),\prod_{j=1}^M
    \Phi_{(\frac{2p}{2p+1},\frac{2p}{2p+1}|\frac{2p}{2p+1},
    \frac{2p}{2p+1})}(w_j,\bar w_j)\prod_{i=1}^N
    \Phi_{(0,0|1,1)}(z_i,\bar z_i)\Omega_0\rangle=}
    \nonumber\\
    & & \prod_{j<j'}|w_i-w_{j'}|^{1/(2p+1)}\prod_{i,j}(z_i-w_j)
    \prod_{i<i'}(z_i-z_{i'})^{2p+1}    
    e^{-\frac{1}{2}\sum_i|z_i|^2-
    \frac{1}{2(2p+1)}\sum_j|w_i|^2}\,. 
  \end{eqnarray}
Indeed, the flux quanta have fractional statistics parameter $\theta/\pi 
= 1/(2p+1)$, and fractional charge $-e/(2p+1)$. Thus, they behave as 
anyons \cite{Wil82,MR91}. In this way, we reproduce the basic excitations
of the Laughlin wave functions. Of course, the anti-holomorphic part
$\prod_{j<j'}(\bar w_j - \bar w_{j'})^{1/2(2p+1)}$ drops out by chiral
projection but cannot be avoided due to mathematical consistency: 
The Berry connexion, which actually yields the anyonic statistics, is entirely
determined by the on $w_i,\bar w_i$ dependent normalization of the wave 
function, see \cite{ASW84} or the introduction to chapter 2 in \cite{Sto92}.
This is one reason why we have to work with the full RCFT. Note that
the leading terms of correlators of primary fields are identical to the
correlators of basic vertex operators, if no screening charges are
needed. This especially is the case in (\ref{eq:myvev}).
  \par
We will now read Jain's idea \cite{Jai89a,Jai89b} (but see also \cite{HLR93})
of ``attaching of flux quanta'' literally.
Thus, we let approach the coordinates $w_i\leftarrow z_i$, where for
simplicity we first set $M=N$ and insert the operator product expansion 
(OPE) of $\Phi_{(0,0|1,1)}(z,\bar z)
\Phi_{(\frac{2p}{2p+1},\frac{2p}{2p+1}|\frac{2p}{2p+1},\frac{2p}{2p+1})}
(w,\bar w)$. We would like to remark that OPE can mathematically 
rigorously be defined only for {\em local\/} fields. This further forces us 
to consider the full RCFT with left and right chiral parts.
The OPE takes the general form
  \begin{eqnarray}\lefteqn{%
    \Phi_{(\alpha|\beta)}(z,\bar z)\Phi_{(\gamma|\delta)}(w,\bar w) 
    =\nonumber}\\
    & &\sum_{\zeta,\eta}(z-w)^{h(\zeta)-h(\alpha)-h(\gamma)}
    (\bar z - \bar w)^{\bar h(\eta)-\bar h(\beta)-\bar h(\delta)}
    {\cal C}_{\alpha\gamma}^{\zeta} \bar{{\cal C}}_{\beta\delta}^{\eta}
    \tilde{\Phi}_{(\zeta|\eta)}(w,\bar w)\,,
  \end{eqnarray}
where $\tilde{\Phi}_{(\zeta|\eta)}$ denotes a generic descendant field 
$f(\del\phi,\del^2\phi\ldots)\Phi_{(\zeta|\eta)}$. The fusion rules of 
our RCFT tell us which ${\cal W}$-conformal families will contribute to the 
right hand side of the OPE. It is sufficient to restrict ourselves to the
term of leading order, since we are interested in the mesoscopic effects of
attaching flux quanta. This is sometimes called the fusion product. Taking 
into account the $\BZ_2$ structure due to the character splitting according to
$(-)^F$ and the twisted boundary conditions we have
  \bea\lefteqn{
  \Phi_{(0,0|1,1)}\star
  \Phi_{(\frac{2p}{2p+1},\frac{2p}{2p+1}|\frac{1}{2p+1},\frac{1}{2p+1})}\star
  \Phi_{(\frac{2p}{2p+1},\frac{2p}{2p+1}|\frac{1}{2p+1},\frac{1}{2p+1})}}
  \nonumber\\
  &=&
  \Phi_{(\frac{2p+2}{2p+3},-\frac{2p+2}{2p+3}|\frac{1}{2p+1},\frac{1}{2p+1})}
  \star\Phi_{(\frac{2p}{2p+1},\frac{2p}{2p+1}|\frac{1}{2p+1},
  \frac{1}{2p+1})} + \ldots \nonumber\\
  &=&
  \Phi_{(0,0|1,1)}\star
  \Phi_{(\frac{p+1}{2p+3},-\frac{p+1}{2p+3}|\frac{2p+2}{2p+3},
  -\frac{2p+2}{2p+3})} + \ldots \nonumber\\
  &=& 
  \Phi_{(\frac{1}{2p+3},-\frac{1}{2p+3}|\frac{2p+2}{2p+3},
  -\frac{2p+2}{2p+3})} + \ldots \,.
  \eea
We see that the leading term of the OPE is associative, but locality is only
preserved, if fusion is done twice. This is expected, since we work with
frustrated boundary conditions and the partition function $Z_{1,0}$ is not
invariant under the $S$-transformation. Thus, attaching {\em two\/} flux
quanta is a well defined operation within our RCFT, but attaching only one is
not, since it destroys locality.
It is surprising that attaching two flux quanta changes the statistics of
the system from $\theta/\pi=1/(2p+1)$ to $\theta'/\pi=1/(2p+3)$. Moreover,
since the total magnetic flux density must remain constant 
\cite{Jai89a,Jai89b,HLR93},
the size of the system must change. This dissipation of the system (since we 
must decrease the electron density) will cost energy to compensate the 
pressure of the external magnetic field, and therefore, cool down the new
QHE state. But the underlying statistics has changed such that the
state cannot go to the old groundstate. On the other hand, changing the size
of the system also redefines the periodicity conditions of the underlying free
field (due to a shift in the magnetic length), and consequently the energy
scale provided by the zero mode of the Virasoro field. The central term
$\frac{c}{24}$ changes by a shift of $-1$, but the energy of the system
renormalizes by a shift of $\frac{1}{4p+6}$ in each chiral sector, the
smallest possible amount admitted by the spectrum of the RCFT.
Taking both effects into account
we exactly end up with the groundstate $\Phi_{(0,0|1,1)}$ of the RCFT with
central charge $c=1-12(2p+3)$. Introducing a further label for the
square of the background charge $\alpha_0$, we have a map 
  \be\label{eq:map}
    F:\ \Phi_{(0,0|1,1)}^{(\frac{2p+1}{2})}
    \stackrel{2\times{\rm fusing}}{\longmapsto}
    \Phi_{(\frac{1}{2p+3},-\frac{1}{2p+3}|\frac{2p+2}{2p+3},
    -\frac{2p+2}{2p+3})}^{(\frac{2p+1}{2})} 
    \stackrel{\rm resizing}{\longmapsto}
    \Phi_{(0,0|1,1)}^{(\frac{2p+3}{2})}\,.
  \ee
  \par
The theory with $c=1-12(2p+1)$ has two statistics sectors with
$\theta=\pi/(2p+1)$ and $\theta'=\pi/(2p+3)$. The latter is -- up to an
overall shift in the conformal weights -- identical to the first sector of
the theory with $c=1-12(2p+3)$. 
  \par
But how can we really change from one RCFT to another? Experimentally
\cite{KHKP}, one
observes sharp phase transitions between the Hall plateaus with universal
critical exponent $7/3$. The problem is that the transition between Hall
plateaus also changes the scale in the system, the magnetic length. In our
present picture we further need to bridge over a gap $\Delta c=-24$ in the
central charge. We propose now that this can be done with the help of
still another CFT, which contains scale dependent logarithmic operators.
Such theories can be consistently defined \cite{Gur93} and can even be
rational \cite{Flo95}. Actually, all theories with $c=c_{p,1}=13-6(p+p^{-1})$ 
share this property. And there is one extremely interesting candidate:
The logarithmic RCFT with $c=c_{6,1}=-24$ contains a primary field 
$\Phi_{2,3}$ with conformal weight $h_{2,3}=7/3$. Therefore, we
conjecture that resizing the system by changing the external magnetic field
can be described by tensoring our first RCFT with the $c_{6,1}$ 
model. The second arrow in our ``fusion map'' (\ref{eq:map}) would 
then provided by 
  \be\label{eq:fmap}
    F:\ {\euf C}\left[{\phantom{2p}1\phantom{+\ 2p}1\phantom{+}\atop
    2p+1\ 2p+3}\right]\otimes{\euf C}[c_{6,1}] \mapsto
    {\euf C}\left[{\phantom{2p}1\phantom{+\ 2p}1\phantom{+}\atop
    2p+3\ 2p+5}\right]\otimes{\euf C}[c_{3,2}]\,, 
  \ee
where we also need the non-unitary $c_{3,2}=0$ theory to get the effective 
central
charges matched. As is known, such non-unitary CFTs with $c=0$ can describe
(de-)localization effects due to disorder, which presumably are
necessary for (un-)bounding flux quanta, see \cite{Zir94} and references 
therein. To be more specific,
tensoring with the field $\Phi_{0,1}^{(c_{1,6})}$ of conformal weight
$h_{0,1}=-1$ maps the second sector of one RCFT to the first of the 
``next'' theory, 
\be  
  h_{\frac{\lambda}{2p+3},\frac{-\lambda}{2p+3}}^{(c=1-
  12(2p+1))} + h_{r,s}^{(c_{6,1})} = h_{\frac{\lambda}{2p+3},
  \frac{\lambda}{2p+3}}^{(c=1-12(2p+3))}+h_{r',s'}^{(c_{3,2})}
\ee 
with $3r'-2s' = 6r-s$. The filling factor changes by this operation as
$\nu^* = ST^2S(\nu) = \frac{\nu}{2\nu+1}$. With this we can match the
central charge as well as its effective value, but not the entire 
spectrum of the RCFTs. This can be done by explicitly introducing disorder
to the system. Assume that the disorder can be described by a random vector
potential $(A_z,A_{\bar z})$ with Gaussian measure
  $$
    P[A]=\exp\left(-\frac{1}{\sigma}\int\frac{d^2 x}{\pi}A_zA_{\bar z}\right)
    \,,
  $$
which enters the action by an additional term
  $$
    S_{{\em disorder}} = i\int\frac{d^2 x}{\pi}\left(
    A_{\bar z}\del\phi + A_z\bar{\del}\phi\right)\,.
  $$
Since our model is Gaussian, the disorder factorizes (see e.g.\ \cite{Ber95}).
Moreover, each
of the two sectors of our RCFT can be affected separately by the disorder,
i.e.\ we may introduce two different disorder couplings $\sigma,\sigma'$.
The conformal dimensions are affected as
  \be
  h^{{\em disorder}}_{\frac{\lam}{2k+\eps},(-)^{\eps}\frac{\lam}{2k+\eps}}
  = h^{{\em pure}}_{\frac{\lam}{2k+\eps},(-)^{\eps}\frac{\lam}{2k+\eps}}
  - \sigma\frac{\lam^2}{4k+2\eps}\,. 
  \ee
We denote the disordered theory
by ${\euf C}_{\sigma,\sigma'}\left[{p'\ p\atop q\ q'}\right]$.
It is easy to check that the spectra of both sides of (\ref{eq:fmap})
can be matched for various choices of the disorder couplings. 
One particular simple choice is
  \be\label{eq:fdmap}
    {\euf C}_{\sigma,0}
    \left[{\phantom{2p}1\phantom{+\ 2p}1\phantom{+}\atop
    2p+1\ 2p+3}\right]\otimes{\euf C}[c_{6,1}] =
    {\euf C}_{0,\sigma'=\frac{(2p+5)\sigma-4}{2p+1}}
    \left[{\phantom{2p}1\phantom{+\ 2p}1\phantom{+}\atop
    2p+3\ 2p+5}\right]\otimes{\euf C}[c_{3,2}]\,, 
  \ee
with fixpoint $\sigma'=\sigma$ for $\sigma=1$. With this choice the
disorder only affects part of the states. Moreover, there is a
minimal disorder such that the spectra can match, since the disorder
coupling $\sigma\geq 0$. It is $\sigma=\frac{4}{2p+5}$. This may provide
an argument for the stability of the plateaus. The details of this 
construction will be presented elsewhere \cite{Flo??}. In the following
we will omit the explicit notion of disorder and will understand
equations of tensor products of RCFTs as disorder driven transitions
{}from the left hand side to the right hand side.
%
%
  \par
  \mysection{Landau Levels of Composite Fermions}
  \pn 
Jain obtains other filling factors of the form $\nu = \frac{n}{2pn\pm 1}$
by filling $n$ Landau levels of composite fermions with $2p$ attached
fluxes. The usual description of QHE states with filling $\nu = n_e/n_f$
within the CFT picture involves theories with $c_{{\em eff}} = n_e$. Within 
this scheme
it is difficult to project down to the lowest Landau level which usually
is done for calculation of the wave functions, and for their comparison with
numerical studies. Moreover, recently
it has been shown \cite{CTZ95} that QHE states can be given as 
lowest-weight representations of minimal models of the $\w_{1+\infty}$ 
symmetry algebra of incompressible quantum droplets found earlier in 
\cite{CTZ93,FlVa93,Kar94}. This explicitly reduces the symmetry down to only
one (charged) $\hat{\euf u}(1)$ current and a classical ${\euf su}(n)$
symmetry of neutral excitations.
  \par
Therefore, one should divide out additional $\hat{\euf u}(1)$ currents
{}from the RCFT. Consider the case $n=2$, i.e.\ we have to tensorize
two theories in our scheme.
For example, the theory 
${\euf C}\left[{\phantom{2p}1\phantom{+\ 2p}1\phantom{+}\atop
2p+1\ 2p+3}\right]\otimes
{\euf C}\left[{\phantom{2p}1\phantom{+\ 2p}1\phantom{+}\atop
2p+1\ 2p+3}\right]/\hat{\euf u}(1)$ 
has central charge $c=(1-12(2p+1))+(1-12(2p+1))-1 = 1 - 24(2p+1)$ 
and effective central charge $c_{{\em eff}} = 1 + 1 - 1 = 1$, hence must be
one of the possible bosonic non-unitary RCFTs with $c_{{\em eff}} = 1$,
say ${\euf C}\left[{\tilde p\ p\atop q\ \tilde q}\right]$.
But counting the number of flux quanta, we find a missing defect of
$2p$, precisely because the tensor product alone does not incorporate the
interaction of fluxes of composite fermions of one Landau level with the
charges of the particles in another. Thus, in order to provide the
missing flux quanta, we have to tensorize again with 
${\euf C}[c_{6,1}]^{\otimes 2p}$. Then we finally end up with a bosonic
RCFT of central charge $c = 1 - 24(2p+1) - 24(2p) = 1 - 12\cdot 2(4p+1)$,
modulo $c=0$ contributions, in which we hide all disorder effects.
  \par
If the filling factor is $\nu=n_e/n_f$, then the corresponding RCFT has central
charge $c=1 - 12n_en_f$. The fact that the square of the background charge can
be factorized $2\alpha_0^2=n_en_f$ with $(n_e,n_f)=1$ means that the
(unfrustrated) partition function of the corresponding RCFT is non-diagonal. 
As a consequence, there may exist a primary field $\Phi_{h,\bar h}(z,\bar z)$
with weights $(h,\bar h)$ such that $h+\bar h\leq h_{{\em min}}$. Then this
field serves as groundstate in the frustrated theory with partition
function $Z_{1,0}$ and should therefore be used for building the wave
functions.
Note, that dividing out the current at the same time yields the projection
to the lowest Landau level, since the effective number of degrees of
freedom reduces to $c_{{\em eff}} = 1$. 
  \par
The wave function for a QHE state with $\nu = 2/(4p+1)$ is given by the
following expression of correlation functions of the fermionic theory
${\euf C}\left[{\phantom{2p}1\phantom{+\ 2p}1\phantom{+}\atop 
2p+1\ 2p+3}\right]$, with $c = 1 - 12(2p+1)$ and $\BZ_2$-twisted boundary 
conditions, where we denote only the left-chiral part for simplicity:
  \bea\label{eq:2LL}
    \Psi_{\nu=\frac{2}{4p+1}}(\{z,w\})
    &=&
    \frac{
      \left\langle
        \prod_{i=1}^{N}\Phi_{0,0}(z_i)
        \prod_{j=1}^{N}\Phi_{\frac{1}{2p+1},\frac{1}{2p+1}}(w_j)
      \right\rangle
      \left\langle
        \prod_{i=1}^{N}\Phi_{0,0}(w_i)
        \prod_{j=1}^{N}\Phi_{\frac{1}{2p+1},\frac{1}{2p+1}}(z_j)
      \right\rangle
    }{
      \left\langle
        \prod_{i=1}^{N}\Phi_{\frac{1}{2p+1},\frac{1}{2p+1}}(z_i)
      \right\rangle
      \left\langle
        \prod_{i=1}^{N}\Phi_{\frac{1}{2p+1},\frac{1}{2p+1}}(w_i)
      \right\rangle
    }\nonumber\\
    &=&
    \prod_{i<j}(z_i-z_j)^{2p+1}\prod_{i<j}(w_i-w_j)^{2p+1}
    \prod_{i,j}(z_i-w_j)^{2p}\,.
  \eea
This expression is just the interaction of composite fermions of one Landau
level with $2p$ flux quanta of the composite fermions of the other Landau
level times vice versa and self-interaction of the flux quanta divided out.
In fact, the primary field
$\Phi_{\frac{1}{2p+1},\frac{1}{2p+1}}(z)=$ 
:$(\Phi_{\frac{2p}{2p+1},\frac{2p}{2p+1}})^{2p}$:$(z)$ 
is the leading term of the $2p$-fold normal ordered product of vertex operators
describing single flux quanta. The result is identical to the wave functions
proposed by Chern-Simons theory. 
  \par
There are now two ways to obtain wave functions in the lowest Landau level:
The first is to describe particles from Landau level $n$ by insertion of
fields $\partial^{n-1}\Phi_{0,0}(z_i)$. Thus the (unnormalized) wave function 
for a state with $n$ filled Landau levels would be
  \be\label{eq:nLL}
    \Psi_{\nu=\frac{n}{2pn+1}}(\{z^{(0)},z^{(1)},\ldots,z^{(n-1)}\}) =
    \left\langle
      \prod_{i=1}^{N/n}\Phi_{0,0}(z^{(0)}_i)
      \prod_{j=1}^{N/n}\partial\Phi_{0,0}(z^{(1)}_j)\ldots
      \prod_{k=1}^{N/n}\partial^{n-1}\Phi_{0,0}(z^{(n-1)}_k)
    \right\rangle\,.
  \ee
For $p=0$ this yields the correct result of the Slater determinant for $n$
Landau levels (after left normal ordering of all $\bar z_i$ variables and
their replacement by $2\partial_j$ to map the function back to Bargmann 
space). But for $p>0$ this ansatz may be too simple.
The second way admits non-analytic powers in the wave function. We use the
RCFT with $c=1-12n(2pn+1)$ and there the field $\Phi_{\frac{n-1}{n},
\frac{n-1}{n}}(z)$, which has minimal conformal dimension in the frustrated
model with non-diagonal partition function according to the factorization
$(1-c)/12 = (n)\cdot(2pn+1)$. 
The (unnormalized) wave function then gets the general form
  \be\label{eq:nuLL}
    \Psi_{\nu}=
    \langle\prod_{i=1}^{\nu N}\Phi_{\frac{n-1}{n},\frac{n-1}{n}}(z_i)\rangle=
    \prod_{1\leq i<j\leq\nu N}(z_i-z_j)^{1/\nu}\,.
  \ee
We may extract the holomorphic part $\prod_{i<j}(z_i-z_j)^{[1/\nu]}$ 
{}from the wave
function (\ref{eq:nuLL}), which for the Jain series is given by $[1/\nu]
= 2p$, i.e.\ a pure Jastrow factor. The remaining non-analytic part lives 
on an $n$-fold covering of the
complex plane which reflects the number of Landau levels. Expanding it in
such a way that the overall asymmetry of the wavefunction is assured and
(after left normal ordering of all still fractional powers modulo one)
replacing $z_i^{a/n}\mapsto(2\partial_i)^a$ for $0<a<n$ yields the wave 
function
projected down to the lowest Landau level. If one skips all negative powers
in the expansion of the non-analytic part, one exactly recovers the Jain
wave functions for $\nu=\frac{n}{2pn\pm 1}$, i.e.\ a Slater determinant of 
$n$ Landau levels multiplied with a Jastrow factor\footnote{The filling
fractions $\frac{n}{2pn-1}$ are obtained by formally exchanging the r{\^o}le
of $z_i$ and $2\partial_i$ in the Slater part of the wave function, i.e.\
setting $\Psi_{\nu=-n}(\{z\}) = \Psi_{\nu=n}(\{\bar z\})$.}. Keeping
the negative powers (where also the r{\^o}le of $z_i$ and $2\partial_i$
has to be exchanged) yields additional terms, which can be viewed as
higher order corrections of the Slater-Jastrow wave functions.
  \par
Note that formally one has two choices for $[1/\nu]$, $1/\nu\not\in\BZ$,
namely $\lfloor 1/\nu\rfloor$ and $\lceil 1/\nu\rceil$, which are not 
equivalent
on the level of our formal expansion of the non-analytic remainder, where
the overall asymmetry of the wave function has to be kept.
This may be understood as a mixing of the two polarization possibilities of
the particles. The partially polarized states are just given by the sum of
the two possible wave functions.
This is supported by the observation that the mixing is exactly
half-half for the states with filling $\nu=2/d$ which are known to be
unpolarized, and one-zero for the Laughlin states which
are known to be fully polarized (here $1/\nu$ is an integer).
In fact, $\Psi_{\nu=\frac{2}{4p\pm 1}}$ can be splitted as
$\Psi_{\nu=\frac{1}{2p}}\Psi_{\nu=\pm 2}$ or as
$\Psi_{\nu=\frac{1}{2p\pm 1}}\Psi_{\nu=\mp 2}$.
  \par
The RCFTs we use in our second approach are the fermionic theories
${\euf C}\left[{r\phantom{\ 2p+1}n\atop
2pn+1\ s}\right]$ with $n(2pn+1)-rs=\pm 2$ for $n$ odd, and the bosonic 
theories ${\euf C}\left[{r\phantom{\ 2p+1}n/2\atop
2pn+1\ \phantom{/2}s}\right]$ with $n(2pn+1)/2-rs=\pm 1$ for $n$ even.
Filling additional Landau levels of composite fermions corresponds to
  \begin{eqnarray}\label{eq:tensor}\lefteqn{%
    T:\ {\euf C}\left[{r\phantom{\ 2p+1}n\atop 2pn+1\ s}\right]\otimes
    {\euf C}\left[{r'\phantom{\ 2p+1}m\atop 2pm+1\ s'}\right]\otimes
    {\euf C}[c_{6,1}]^{\otimes 2pmn}/\hat{\euf u}(1)=\nonumber}\\ 
    & &{\euf C}\left[{\tilde r/2\phantom{\ 2p+1}(n+m)/2\atop 
    2p(n+m)+1\ \phantom{(n/}\tilde s}\right]\otimes
    {\euf C}[c_{3,2}]^{\otimes 2pmn}\,,
  \end{eqnarray}
for $n,m$ odd and similar for $n$ or $m$ even where bosonic and fermionic
theories have to be replaced accordingly. The filling factor transforms due
to this operation as 
$\nu^*=\frac{n+1}{2p(n+1)+1}=(ST^{2p}S)T^m(ST^{-2p}S)(\nu)$. 
Note that the modular
transformation on $\nu$ for filling composite particle Landau levels is
quite complicated and does depend on the number $2p$ of flux quanta already
attached. This is reflected in the fact that our operation in the space
of RCFTs involves a $2pmn$-fold tensor product. Actually, the correct number
of additional flux quanta is not entirely fixed by the condition that the
number of Landau levels is strictly additive. We will show later that
this can explain the appearance of some observed non-Jain series. 
  \par
But first we give a naive argument on the number of additional flux
quanta. Consider two QHE states with $\nu=n_e/n_f$ and $\nu'=n'_e/n'_f$, where
without loss of generality $n_f = 2p_f +1$, $n'_f = 2p'_f + 1$. Joining
these two states in the manner described above to get a new one, we 
introduce interaction between the charges $n_e$ and the fluxes $2p'_f$ and
vice versa, which all together results in $p_fn'_e+p'_fn_e$ fusion
operations to get $\nu^*=\frac{n_e+n'_e}{2p_f+2p'_f+1}$. In fact, 
we not only tensorize the corresponding theories, but contract within
the correlation functions in all possible ways fields from one
theory with fields from the other via fusion $F$. If $\hat{\otimes}$ denotes
tensoring with dividing out $\hat{{\euf u}}(1)$, and if $p_fn'_e = 
p'_fn_e$, we have our second map
  \be
  T:\ F^{p'_fn_e}{\euf C}_{\nu=\frac{n_e}{2p_f+1}}\otimes 
      F^{p_fn'_e}{\euf C}_{\nu'=\frac{n'_e}{2p'_f+1}} 
      \stackrel{\rm joining}{\longmapsto}
      F^{p_fn'_e+p'_fn_e}({\euf C}_{\nu=\frac{n_e}{2p_f+1}}
      \hat{\otimes} 
      {\euf C}_{\nu'=\frac{n'_e}{2p'_f+1}}) 
      \stackrel{\rm resizing}{\longmapsto}
      {\euf C}_{\nu^*=\frac{n_e+n'_e}{2p_f+2p'_f+1}}\,.
  \ee
  \par
It is easy to see that the filling fractions of the Jain series are precisely
the first order of continued fraction expansions 
  \begin{equation}
    \nu   = [n_1,2p_1,\ldots,n_{k+1}]
          = n_{k+1} + \frac{1}{{\ds 2p_k + 
                       \frac{1}{{\ds\ddots +
                        \frac{1}{{\ds n_2 + 
                         \frac{1}{{\ds 2p_1 + \frac{1}{n_1}
                      }} }} }} }}\,.
  \end{equation}
As has been shown in \cite{FrKe91,FZ91}, the QHE can be described
by a Chern-Simons theory. Then, due to large scale principles, only the 
one-loop diagrams contribute to the Hall conductivity. Viewing Chern-Simons 
theory as massive QED in three dimensions, where two flux quanta serve as
a massive photon, this corresponds to a $1/N$ expansion \cite{Flo94up},
which naturally yields the filling factor as a continued fraction expansion
and the Jain series as first order approximation to it.
%
%
  \par
  \mysection{Exceptional cases}
  \pn 
The moduli space of the non-unitary RCFTs with $c_{{\em eff}} = 1$ has been
explored in \cite{Flo93}. It is closely related to the modular group
$\Gamma={\rm PSL}(2,\BZ)$ for $\Delta=1$, and to $\Gamma(2)$ for $\Delta=2$.
So far, we described transitions, which change $\nu$ to first order in its
continued fraction expansion, by our $F$ (fusion) and $T$ (tensoring) 
operation on RCFTs.
Theoretically, every arbitrary filling factor can be obtained in this way.
The $F$-map yields the $ST^{-2}S$
move of the modular group in our moduli space. Tensoring of theories
yields the $T$ move (which goes from the
moduli space of fermionic theories to the space of bosonic theories and
vice versa). Both together implement $\Gamma_T(2)$, one of the possible
three subgroups $\Gamma(2)\subset\Gamma_X(2)\subset\Gamma$ of index two.
  \par
The appearance of a subgroup of the modular group in the QHE has already 
been noted earlier \cite{Kie91,Lut93,LuRo93}. Our approach now yields a new
description of the modular group as directly acting on a moduli space
of RCFTs by the operations of $F$ and $T$. More specifically
we have for example
  \be\label{eq:2tensor}
    T:\ {\euf C}\left[{\phantom{2p}1\phantom{+\ 2p}1\phantom{+}\atop
    2p+1\ 2p+3}\right]\otimes
    {\euf C}\left[{\phantom{2p}1\phantom{+\ 2p}1\phantom{+}\atop 
    2p+1\ 2p+3}\right]\otimes
    {\euf C}[c_{6,1}]^{\otimes 2p}/\hat{\euf u}(1) =
    {\euf C}\left[{\phantom{4p}2\phantom{+\ 2p}1\phantom{+}\atop 
    4p+1\ 2p+1}\right]\otimes
    {\euf C}[c_{3,2}]^{\otimes 2p}\,,
  \ee
which describes the transitions $T(\nu=1/(2p+1),\nu'=1/(2p+1)) 
\mapsto \nu^*=2/(4p+1)$, 
i.e.\ $1\mapsto 2,1/3\mapsto 2/5,1/5\mapsto 2/9,\ldots$ As mentioned above, 
the partition function of the resulting theory is now  partly non-diagonal, 
i.e.\ characters to 
different lowest-weight representations are combined, forcing that in the 
frustrated $Z_{1,0}$ model we have a non-trivial field creating the
groundstate and thus yielding the wave function as its $n$-point function.
Due to the second filled Landau level this correlator is no longer purely
holomorphic but must be expanded in a specific way.
This may be repeated in the following way: Starting with $\nu=n/(2pn+1)$
we have a transition to $\nu^*=(n+1)/(2p(n+1)+1)$ via the RCFT tensor
product (\ref{eq:tensor}) with $m=1$. Together with the fusion description of 
attaching flux quanta this yields all the Jain series -- and more!
  \par
Consider for example the resulting theory in (\ref{eq:2tensor}). Applying
the fusion map we have a transition $F:\ \nu=2/(4p+1)\mapsto\nu^*=4/(2p+1)$.
Form this we naturally obtain the filling fractions $4/5,4/11$ and $4/13$ 
which are not members of the Jain hierarchy to first order\footnote{The
so called particle-hole duality $\nu\leftrightarrow 1-\nu$ (or $2-\nu$)
is not considered in this work, since it is not supported by experiment
in a sufficient way, i.e.\ for many QHE states the particle-hole
duality conjugate state is not observed.}. Also, $5/7$ is obtained by
the fusion map from $3/11=3/(2\cdot 2\cdot 3-1)$. Note that the
interpretation of the fusion map as purely attaching of flux quanta is
no longer valid, if it is applied to QHE states of $n>1$ Landau levels.
Next, $7/11$ is obtained from $3/25=3/(2\cdot 4\cdot 3+1)$, $8/11$ from
$3/29=3/(2\cdot 5\cdot 3-1)$, states with
$p\geq 4$ flux quanta attached, which presumably is extremely difficult to
observe, as is explained below. We propose that in such cases, where the
state obtained from the fusion map has a significantly smaller number $n_f$
of total flux quanta than the original state, the former is realized instead
of the latter.
  \par
We even can get two whole new series out of our tensor operation
(\ref{eq:tensor}) in the following way: 
The number of Landau levels must be additive, i.e.\ $(m+n)$ must be factored
out from the rhs. This is possible for $p=1$ even for a smaller power of
$c_{6,1}$ theories. Namely, we could choose the powers $2(mn-m-n)$ and 
$2(mn-2m-2n)$ to tensorize with 
${\euf C}\left[{\phantom{2p}r\phantom{-1}\ m\atop 2pm-1\ s}\right]\otimes
{\euf C}\left[{\phantom{2p}r'\phantom{-1}\ n\atop 2pm-1\ s'}\right]$
which yields $\nu^*=n/(2n-3)$ and $\nu^*=n/(2n-5)$ respective.
Note that lower members of the $n/(2n-5)$ series cannot be realized,
because negative powers in the tensor product are meaningless, in
particular if we use the physical case $m=1$, i.e.\ addition of just
one Landau level. Therefore, the first member of this series is
the fraction $8/11$.
Other cases such as $p>1$ or a series $n/(2n-7)$ do not yield any 
new fraction in the ``observable region'' of the $(n_e,n_f)$-plane\footnote{
The observable region of the $(n_e,n_f)$-plane for $\nu<1$ is more or
less defined by $n_e\leq 10$, $n_f<20$ and $(n_f - 1)/2<4$, conditions which
stem from phenomenological considerations on the stability of QHE states.}. 
Moreover, since it is physically unlikely that states realize, where the
power of $c_{6,1}$ models had to be higher then $2mn$ (which always 
is satisfactory to get additivity of Landau levels), one does indeed not
observe states with $\nu=n/(2pn+k)$, $k=3,5,\ldots$.
Note that we obtain all observed fractions within our scheme on the
first level, i.e.\ by applying our fusion and tensor map each only once.
  \par
The huge amount of other possible transitions allowed by $\Gamma_T(2)$
all correspond to a change of $\nu$ to higher order in its continued
fraction expansion. There are indeed experimentally observed filling
fractions, which are not of first order, i.e.\ not members of the Jain series.
But all these are of second order. They presumably belong to system
configurations, where the number of particles is small enough such that
second order effects become visible. All of them can be explained by
the exceptional solutions described above. 
  \par
The transitions between QHE plateaus presumably are disorder
driven. In our treatment we found a minimal disorder coupling such that
a transition can take place, which essentially is proportional as
$\sigma\sim\frac{1}{n_en_f}$. 
{}From this it follows that with increasing $n,p$ in the Jain series
$\nu=\frac{n}{2pn\pm 1}$ the plateau width decreases. In particular, it
decreases fast for increasing $p$. Thus, higher members of the Jain
series are more difficult to observe, since the parameters of the
experiment have to be controlled with higher accuracy. 
It can even happen that second 
order states such as for $\nu=4/(2p+1)$ can be seen without 
the corresponding first order states, here $\nu=2/(4p+1)$ for $p>3$,
being observed, since the latter are much less stable against disorder.
Also, if the total number of bound flux quanta $n_f$ is very high, it
seems likely that the system chooses a new configuration with lower $n_f$.
In some cases this is possible via the fusion map which then in fact
unbounds flux, e.g.\ $F:\ \frac{2}{4p+1}\mapsto\frac{4}{2p+1}$ as mentioned
above. We explicitly checked that such possibilities are very rare and do
not produce any unobserved filling fractions (in the sense that $n_e,n_f$
are sufficiently small to yield states which should in principle
be accessible by experiment).
  \par
To conclude, we think that our proposed way of describing transitions 
between QHE states by algebraic operations on the space of RCFTs strongly
supports the composite fermion picture of Jain and may explain several
experimentally observed facts. {\em In particular we have a frame in which
precisely the observed fractions can be explained -- and the
completeness of the set of
already observed series.} Our frame naturally produces the
experimentally observed universal critical exponent $7/3$.
Moreover, it once more shows the deep
r{\^o}le the modular group plays in nature. We hope that it may serve
as a starting point for future investigations. A lot of questions remain open:
Since our approach is
heavily based on CFT numerology, it is urging to support it
by a more close connection to first principle treatments of the FQHE. 
It also would be worthwhile to connect it with the classification of
quantum Hall fluids obtained by Fr{\"o}hlich, Studer, and Thiran 
\cite{FST94,FrTh94},
and with topological explanations for the exclusivity of the Jain series
\cite{Var94}. This will be done in our future work \cite{Flo??}. 
  \bigskip\pano
{\bf Acknowledgment:}
I would like to thank J.~Gaite, W.~Nahm, G.~Sierra, 
and especially C.~Gomez and R.~Varnhagen for useful discussions. 
This work has been supported partly by
the Deutsche Forschungsgemeinschaft and partly by the European
Scientific Network n$^{{\rm o}}$ ERB CHRX CT 920069.
%
%
  
  \end{document}